# Self - Dual Supergravity in Seven Dimensions with Reduced Holonomy $G_2$

Hitoshi NISHINO[1] and Subhash RAJPOOT[2]

*Department of Physics & Astronomy*
*California State University*
*1250 Bellflower Boulevard*
*Long Beach, CA 90840*

## Abstract

We present self-dual $N = 2$ supergravity in superspace for Euclidean seven dimensions with the reduced holonomy $G_2 \subset SO(7)$, including all higher-order terms. As its foundation, we first establish $N = 2$ supergravity without self-duality in Euclidean seven dimensions. We next show how the generalized self-duality in terms of octonion structure constants can be consistently imposed on the superspace constraints. We found two self-dual $N = 2$ supergravity theories possible in 7D, depending on the representations of the two spinor charges of $N = 2$. The first formulation has both of the two spinor charges in the **7** of $G_2$ with $24 + 24$ on-shell degrees of freedom. The second formulation has both charges in the **1** of $G_2$ with $16 + 16$ on-shell degrees of freedom.

---

[1]E-Mail: hnishino@csulb.edu

[2]E-Mail: rajpoot@csulb.edu

1. **Introduction**

It has been well-known that M-theory [1] can produce realistic four-dimensional (4D) theory with chiral fermions upon a particular compactification with extra seven dimensions (7D) with the reduced holonomy $G_2$ [2][3][4] instead of the maximal one $SO(7)$. The reduced holonomy $G_2$ is a special case of a series of reduced holonomies, such as 8D with $Spin(7)$ holonomy, or $G_2$, $SU(3)$ and $SU(2)$ holonomies in 7D, 6D and 4D [5][2][3][4]. In the cases of $Spin(7)$ and $G_2$ holonomies, so-called octonion structure constants play a crucial role [6]. This is because for these reduced holonomies, generalized self-duality conditions can be dictated by octonion structure constants [6] that were not present in the case of self-dual supergravities in 4D [7][8]. In particular, the 7D manifold both with $G_2$ holonomy and generalized self-duality [3][4][5] in the compactification of 11D supergravity is compatible with local supersymmetry, as confirmed by Killing spinors as the singlets of $G_2$ [3][4].

These developments indicate the importance of constructing self-dual supergravity theories with reduced holonomies on these manifolds, as the next natural step to take. Actually, in our previous paper [9], we have carried out such a construction of self-dual supergravity in 8D with reduced holonomy $Spin(7) \subset SO(8)$. We have found that self-dual supergravity in 8D has differences from, as well as similarities to self-dual supergravity in 4D [7][8]. The most fundamental difference is the involvement of octonion structure constants [6] making the whole computation non-trivial.

Recently similar but different formulations have been presented, such as using BRST or topological quantum field symmetry as the guiding principle for constructing self-dual supergravity in 8D or 7D with lower-order terms [10]. However, there seems to be no complete self-dual supergravity formulation in 7D with desirable reduced holonomy $G_2$ [5][6][3] *before* quantization, with all the higher-order interactions in a closed form. For example, in the topological quantum field formulation of self-dual supergravity in [10], only lower-order terms are compared with supergravity theory, due to the complication at fermionic quartic terms. Moreover, topological formulations [10] rely on the BRST symmetry at the quantized level after gauge-fixings, and as such, they are *not* classically gauge-invariant.

In our present paper, we will formulate complete $N = 2$ self-dual supergravity in Euclidean 7D with the reduced holonomy $G_2 \subset SO(7)$. Even though the existence of such a formulation has been conjectured for some time, the required computations for a complete theory are considerably non-trivial, similarly to the case of self-dual supergravity in 8D [9]. The most important objective of this paper is to complete the self-dual supergravity in 7D with the reduced holonomy $G_2$, including all the higher-order interaction terms in superspace, in a self-contained and economical fashion.

As we have done in 8D [9], we adopt a very special set of constraints called 'Beta-Function-Favored Constraints' (BFFC). This set of constraints had been developed for drastically simplifying $\beta$-function computations for Green-Schwarz superstring in 10D [11]. For example, the whole $\beta$-function computation is reduced to the evaluation of only one single Feynman graph [11]. As in the corresponding case in 8D [9], we will see the power of BFFC for simplifying our computations, in particular, for the consistent supersymmetrization of self-duality conditions possible in 7D with the reduced holonomy $G_2$.

Based on this BFFC constraints, we first formulate self-dual supergravity with 'restricted' $N = 2$ supersymmetry with both spinor charges in the **7** of $G_2$. Such a formulation can be given in terms of extra constraints that are superspace generalization of a bosonic generalized self-duality condition $R_{ab}{}^{cd} = (1/2)\phi^{cdef}R_{abef}$ for a Riemann tensor with the (dual) octonion structure constant $\phi^{abcd}$. We next give an alternative self-dual supergravity with 'restricted' $N = 2$ supersymmetry with both spinor charges in the **1** of $G_2$. Due to the 'nilpotent' character of the spinor charges, the latter supergravity can be also regarded as topological gravity, related to the quantum field theories in [10].

## 2. $N = 2$ Supergravity in Euclidean 7D

Before imposing supersymmetric generalized self-duality conditions, we establish first $N = 2$ superspace supergravity in Euclidean 7D with the signature $(+ + \cdots +)$. This process is analogous to self-dual supergravity in 8D [9]. Namely, we use a particular set of constraints BFFC out of infinitely many possible sets of superspace constraints, linked by super-Weyl rescalings [12]. In other words, we establish a 7D analog of the BFFC in 10D [11] or 8D [9]. The BFFC constraints greatly simplifies the whole computation, such as many fermionic terms considerably simplified, or no dilaton in exponents [11][9].

The field content of our $N = 2$ supergravity multiplet is $(e_m{}^a, \psi_m{}^{\underline{\alpha}}, C_{mn}, A_m{}^i, B_m, \chi_{\underline{\alpha}}, \varphi)$ which is formally the same as $N = 2$ supergravity in Minkowskian 7D [13]. The component fields $A_m{}^i$ $(i = 1, 2)$, $B_m$ and $C_{mn}$ have the field strengths $F_{mn}{}^i$, $G_{mn}$ and $H_{mnr}$, respectively. Here we use the *underlined* spinorial indices $\underline{\alpha}, \underline{\beta}, \cdots$ including $N = 2$ indices $A, B, \cdots = 1, 2$, so that $\underline{\alpha} \equiv \alpha A, \underline{\beta} \equiv \beta B, \cdots$, where $\alpha, \beta, \cdots = 1, 2, \cdots, 8$ are for the **8** spinors of $SO(7)$. These indices are also used for fermionic coordinates as usual in superspace [14], while the indices $m, n, \cdots = 1, 2, \cdots, 7$ are for curved bosonic coordinates, and $a, b, \cdots = 1, 2, \cdots, 7$ for local Lorentz bosonic coordinates. Even though superscript/subscript of these bosonic indices does not matter, we sometimes use them to elucidate their contractions. In the Clifford alge-bra for Euclidean 7D, we have a symmetric charge conjugation matrix [15], which can be identified with an unit matrix: $C_{\alpha\beta} = \delta_{\alpha\beta}$. Subsequently, the raising/lowering of spinor indices will not matter, even though we sometimes use their superscripts/subscripts, in order to elucidate contractions. Relevantly, we have the symmetry $(\gamma_{[n]})_{\alpha\beta} = +(-1)^{n(n+1)/2}(\gamma_{[n]})_{\beta\alpha}$ [15], where the symbol $[n]$ implies the totally antisymmetric bosonic indices in order to save space: $\gamma_{[n]} \equiv \gamma_{a_1\cdots a_n}$.

In our superspace, there are superfield strengths $F_{AB}{}^i$, $G_{AB}$, $H_{ABC}$ together with the supertorsion $T_{AB}{}^C$ and supercurvature $R_{ABc}{}^d$, which satisfy the Bianchi identities

$$\tfrac{1}{2}\nabla_{[A}T_{BC)}{}^D - \tfrac{1}{2}T_{[AB|}{}^E T_{E|C)}{}^D - \tfrac{1}{4}R_{[AB|e}{}^f(\mathcal{M}_f{}^e)_{|C)}{}^D \equiv 0 \ , \tag{2.1a}$$

$$\tfrac{1}{6}\nabla_{[A}H_{BCD)} - \tfrac{1}{4}T_{[AB|}{}^E H_{E|CD)} - \tfrac{1}{4}F_{[AB}{}^i F_{CD)}{}^i + \tfrac{1}{4}G_{[AB}G_{CD)} \equiv 0 \ , \tag{2.1b}$$

$$\tfrac{1}{2}\nabla_{[A}F_{BC)}{}^i - \tfrac{1}{2}T_{[AB|}{}^D F_{D|C)}{}^i \equiv 0 \ , \tag{2.1c}$$

$$\tfrac{1}{2}\nabla_{[A}G_{BC)} - \tfrac{1}{2}T_{[AB|}{}^D G_{D|C)} \equiv 0 \ . \tag{2.1d}$$

As has been mentioned, or as in analogous theory in 8D [9], we need to find a BFFC set of constraints. After trials and errors, we found the BFFC set in 7D to be

$$T_{\underline{\alpha\beta}}{}^c = +(\gamma^c \tau_3)_{\underline{\alpha\beta}} \equiv (\gamma^c)_{\alpha\beta}(\tau_3)_{AB} \ , \tag{2.2a}$$

$$T_{\underline{\alpha\beta}}{}^{\underline\gamma} = +i\delta_{\underline{\alpha\beta}}\chi^{\underline\gamma} - i(\gamma^a\tau_3)_{\underline{\alpha\beta}}(\gamma_a\tau_3\chi)^{\underline\gamma} + i(\tau_i)_{\underline{\alpha\beta}}(\tau_i\chi)^{\underline\gamma} - i\delta_{(\underline\alpha}{}^{\underline\gamma}\chi_{\underline\beta)} \ , \tag{2.2b}$$

$$H_{\underline{\alpha\beta}c} = +\tfrac{1}{2}(\gamma_c\tau_3)_{\underline{\alpha\beta}} \ , \quad H_{\underline\alpha bc} = 0 \ , \quad F_{ab}{}^i = 0 \ , \quad G_{\underline\alpha b} = 0 \ , \tag{2.2c}$$

$$F_{\underline{\alpha\beta}}{}^i = -\tfrac{i}{\sqrt{2}}(\tau_i)_{\underline{\alpha\beta}} \ , \quad G_{\underline{\alpha\beta}} = -\tfrac{i}{\sqrt{2}}\delta_{\underline{\alpha\beta}} \ , \quad T_{\underline\alpha b}{}^c = 0 \ , \tag{2.2d}$$

$$T_{\underline\alpha b}{}^{\underline\gamma} = -\tfrac{1}{2}(\gamma^{cd})_{\underline\alpha}{}^{\underline\gamma}H_{bcd} - \tfrac{i}{\sqrt{2}}(\gamma^c\tau_3\tau_i)_{\underline\alpha}{}^{\underline\gamma}F_{bc} - \tfrac{i}{\sqrt{2}}(\gamma^c\tau_3)_{\underline\alpha}{}^{\underline\gamma}G_{bc} \ , \tag{2.2e}$$

$$\nabla_{\underline\alpha}\chi_{\underline\beta} = +\tfrac{i}{\sqrt{2}}(\gamma^c\tau_3)_{\underline{\alpha\beta}}\nabla_c\varphi - \tfrac{i}{12}(\gamma^{[3]}\tau_3)_{\underline{\alpha\beta}}H_{[3]} - \tfrac{i}{4\sqrt{2}}(\gamma^{cd}\tau_i)_{\underline{\alpha\beta}}F_{cd}{}^i + \tfrac{i}{4\sqrt{2}}(\gamma^{cd})_{\underline{\alpha\beta}}G_{cd}$$

$$+ \tfrac{i}{16}(\gamma^a)_{\underline{\alpha\beta}}\chi_a - \tfrac{i}{32}(\gamma^{ab})_{\underline{\alpha\beta}}\chi_{ab} + \tfrac{i}{16}(\gamma^a\tau_i)_{\underline{\alpha\beta}}\chi_{ai}$$

$$- \tfrac{i}{32}(\gamma^{ab}\tau_i)_{\underline{\alpha\beta}}\chi_{abi} - \tfrac{i}{8}(\tau_3)_{\underline{\alpha\beta}}\chi_3 + \tfrac{i}{192}(\gamma^{[3]}\tau_3)_{\underline{\alpha\beta}}\chi_{[3]3} \ , \tag{2.2f}$$

$$\nabla_{\underline\alpha}\varphi = -\tfrac{i}{\sqrt{2}}\chi_{\underline\alpha} \ , \tag{2.2g}$$

$$T_{ab}{}^c = +2H_{ab}{}^c \ , \tag{2.2h}$$

$$R_{\underline{\alpha\beta}cd} = +\sqrt{2}i(\tau_i)_{\underline{\alpha\beta}}F_{cd}{}^i - \sqrt{2}i\delta_{\underline{\alpha\beta}}G_{cd} \ , \tag{2.2i}$$

at the mass dimensions $d \leq 1$. Here $\chi_{[n]} \equiv (\overline\chi\gamma_{[n]}\chi)$, $\chi_3 \equiv (\overline\chi\tau_3\chi)$, $\chi_{[n]i} \equiv (\overline\chi\gamma_{[n]}\tau_i\chi)$, $\chi_{[n]3} \equiv (\overline\chi\gamma_{[n]}\tau_3\chi)$, and the meaning of the underlined indices is, e.g., $(\tau_3)_{\underline{\alpha\beta}} \equiv \delta_{\alpha\beta}(\tau_3)_{AB}$, $\delta_{\underline{\alpha\beta}} \equiv \delta_{\alpha\beta}\delta_{AB}$, while $(\gamma_a\tau_3\chi)^{\underline\gamma} \equiv (\gamma_a\tau_3)^{\underline{\gamma\delta}}\chi_{\underline\delta}$, etc. The $\tau_i$ $(i = 1, 2)$ and $\tau_3$ are

$2 \times 2$ matrices for the $N = 2$ indices:

$$\tau_1 \equiv \begin{pmatrix} 0 & +1 \\ +1 & 0 \end{pmatrix} , \quad \tau_2 \equiv \begin{pmatrix} +1 & 0 \\ 0 & -1 \end{pmatrix} , \quad \tau_3 \equiv \begin{pmatrix} 0 & -1 \\ +1 & 0 \end{pmatrix} . \tag{2.3}$$

There are two important features in these constraints which are similar to the self-dual supergravity in 8D [9]. First, the fermionic components $H_{\underline{\alpha}bc}$, $F_{\underline{\alpha}b}$ and $G_{\underline{\alpha}b}$ are absent, in contrast to any non-BFFC set, where they contain linear dilatino. Second, no exponential factor with the dilaton appears anywhere in our constraints as in 8D [9] or 10D [11]. With its technical details skipped, we mention that the most frequently-used relationship in these computations is the Fierz identity

$$(\gamma_a \tau_3)_{(\underline{\alpha}\underline{\beta}|} (\gamma^a \tau_3)_{|\underline{\gamma})\underline{\delta}} + \delta_{(\underline{\alpha}\underline{\beta}} \delta_{\underline{\gamma})\underline{\delta}} - (\tau_i)_{(\underline{\alpha}\underline{\beta}|} (\tau_i)_{|\underline{\gamma})\underline{\delta}} \equiv 0 . \tag{2.4}$$

Other constraints at $d \geq 3/2$ in our BFFC are

$$\nabla_{\underline{\alpha}} F_{bc}{}^i = -\tfrac{i}{\sqrt{2}} (\tau^i T_{bc})_{\underline{\alpha}} , \tag{2.5a}$$

$$\nabla_{\underline{\alpha}} G_{bc} = -\tfrac{i}{\sqrt{2}} T_{bc\underline{\alpha}} , \tag{2.5b}$$

$$\nabla_{\underline{\alpha}} H_{bcd} = +\tfrac{1}{4} (\gamma_{[b|} \tau_3 T_{|cd]})_{\underline{\alpha}} , \tag{2.5c}$$

$$R_{\underline{\alpha}bcd} = -(\gamma_b \tau_3 T_{cd})_{\underline{\alpha}} , \tag{2.5d}$$

$$\nabla_{\underline{\gamma}} T_{ab}{}^{\underline{\delta}} = -\tfrac{1}{4} (\gamma^{cd})_{\underline{\gamma}}{}^{\underline{\delta}} (R_{cdab} + 2 F_{ab}{}^i F_{cd}{}^i - 2 G_{ab} G_{cd})$$
$$\qquad - \epsilon_{ij} (\tau_3)_{\underline{\gamma}}{}^{\underline{\delta}} F_{ac}{}^i F_{bc}{}^j - (\tau_i)_{\underline{\gamma}}{}^{\underline{\delta}} F_{[a|c}{}^i G_{|b]c}$$
$$\qquad + \tfrac{i}{\sqrt{2}} (\gamma^c \tau_3 \tau_i)_{\underline{\gamma}}{}^{\underline{\delta}} \nabla_c F_{ab}{}^i + \tfrac{i}{\sqrt{2}} (\gamma^c \tau_3)_{\underline{\gamma}}{}^{\underline{\delta}} \nabla_c G_{ab}$$
$$\qquad + i T_{ab\underline{\gamma}} \chi^{\underline{\delta}} - i (\gamma_c \tau_3 T_{ab})_{\underline{\gamma}} (\gamma^c \tau_3 \chi)^{\underline{\delta}}$$
$$\qquad + i (\tau_i T_{ab})_{\underline{\gamma}} (\tau_i \chi)^{\underline{\delta}} - T_{ab}{}^{\underline{\delta}} \chi_{\underline{\gamma}} - i \delta_{\underline{\gamma}}{}^{\underline{\delta}} (\overline{T}_{ab} \chi) , \tag{2.5e}$$

$$R_{[ab]} = +2 \nabla_c H_{ab}{}^c , \tag{2.5f}$$

$$R_{a[bcd]} = -4 \nabla_a H_{bcd} - \tfrac{1}{2} F_{[ab}{}^i F_{cd]}{}^i + \tfrac{1}{2} G_{[ab} G_{cd]} , \tag{2.5g}$$

$$R_{abcd} - R_{cdab} = -2 \nabla_{[a} H_{b]cd} + 4 H_{ab}{}^e H_{ecd} + 4 H_{c[a|}{}^e H_{e|b]d}$$
$$\qquad + 2 F_{ab}{}^i F_{cd}{}^i + 2 F_{a[c}{}^i F_{d]b}{}^i - 2 G_{ab} G_{cd} - 2 G_{a[c} G_{d]b} . \tag{2.5h}$$

These with Bianchi identities at $d = 3/2$ and $d = 2$ lead to the gravitino, graviton, and antisymmetric tensor superfield equations:

$$(\gamma^b T_{ab})_{\underline{\gamma}} - 2i (\tau_3 \nabla_a \chi)_{\underline{\gamma}} + i (\gamma^{bc} \tau_3 \chi)_{\underline{\gamma}} H_{abc} + \sqrt{2} (\gamma^b \tau_i \chi)_{\underline{\gamma}} F_{ab}{}^i + \sqrt{2} (\gamma^b \chi)_{\underline{\gamma}} G_{ab} \doteq 0 , \tag{2.6a}$$

$$R_{ab} + 2 (F_{ac}{}^i F_b{}^{ci} - G_{ac} G_b{}^c) + 2\sqrt{2} \nabla_a \nabla_b \varphi \doteq 0 . \tag{2.6b}$$

$$R_{[ab]} = 2 \nabla_c H_{ab}{}^c \doteq -4\sqrt{2} H_{ab}{}^c \nabla_c \varphi + 2i (\overline{T}_{ab} \chi) . \tag{2.6c}$$

where expressions such as $(\gamma_b T_{cd})_\gamma$ involve the gravitino superfield strength $T_{cd}{}^\delta$, e.g., $(\gamma_b T_{cd})_{\underline\gamma} \equiv (\gamma_b)_{\underline{\gamma\delta}} T_{cd}{}^{\underline\delta}$. The symbol $\doteq$ stands for a superfield equation of motion. Reflecting the Euclidean nature of our 7D, the $F^2$- and $G^2$-terms in (2.6b) have opposite signs, similarly to the corresponding 8D case [9]. As usual in superspace [11][9], the superfield equations for dilatino and dilaton are obtained by the multiplication of (2.6a) by $\gamma^a$ and the trace of (2.6b), respectively.

There are several remarks in order. First, note that (2.5d) corresponds to the supersymmetry transformation of the Lorentz connection $\phi_{bcd}$. In particular, the indices $cd$ are on the gravitino superfield strength $T_{cd}{}^{\underline\beta}$, which is made possible by the particular choice of the bosonic supertorsion component (2.2h), as will be seen in (3.8) and (3.9). Second, similar feature is found in the component $R_{\underline{\alpha\beta}cd}$ (2.2i), where the pair of indices $cd$ appears on the superfield strengths $F_{cd}$ and $G_{cd}$. Third, note the particular order of indices $cdab$ on the Riemann supercurvature $R_{cdab}$ in (2.4e). To reach this form, we made use of the identity (2.5h). Eq. (2.5h) is verified by (2.5g), while the latter is confirmed by the $T$-BI at $d=2$. The first equality in (2.6c) is the same as (2.5f). As in 8D [9], we need the last pair of indices $ab$ in $R_{cdab}$ free in (2.5e), instead of the first pair $cd$, because we can *not* impose the self-duality on the first pair of indices of $R_{cdab}$, but only on the last one, due to the presence of supertorsion $T_{ab}{}^c$.

## 3. $N=2$ Self-Dual Supergravity in 7D with $G_2$ Holonomy

We next present an $N=2$ self-dual supergravity with the reduced holonomy $G_2$. The difference from the last section is that now the spinor charges form the **7** of $G_2$. Interestingly, we can accomplish both the supersymmetrization of the self-duality condition, and the reduction of the maximal holonomy $SO(7)$ into $G_2$ in 7D.

As a first trial of finding a desirable set of superspace constraints, we can try some dimensional reduction from 8D [9] into 7D. However, we soon find that this will not work as smoothly as we first anticipated. In fact, we had a similar experience for the globally supersymmetric self-dual theories in 7D [16]. The main reason is that the reduced holonomy structure $G_2$ complicates such a dimensional reduction. As will be also seen shortly, certain differences in structure of supersymmetries in 7D compared with 8D [9], also cause such complications.

After direct trials and errors within 7D, we have found the following set as the right constraints in addition to the BFFC (2.2) before imposing self-duality:

$$\nabla_{\underline\alpha} \stackrel{*}{=} \mathcal{N}_{\underline{\alpha\beta}} \nabla_{\underline\beta} \equiv (\mathcal{N}\nabla)_{\underline\alpha} \quad , \quad \mathcal{N} \equiv \tfrac{7}{8}\left(I - \tfrac{1}{7}\Psi\right) \quad , \quad \Psi \equiv \tfrac{1}{4!}\phi^{[4]}\gamma_{[4]} \equiv \tfrac{i}{3!}\psi^{[3]}\gamma_{[3]} \quad , \qquad (3.1a)$$

$$T_{ab}^{(-)\underline{\gamma}} \equiv N_{ab}{}^{cd} T_{cd}{}^{\underline{\gamma}} \stackrel{*}{=} 0 \ , \quad N_{ab}{}^{cd} \equiv \tfrac{1}{6}\left(\delta_a{}^{[c}\delta_b{}^{d]} - \phi_{ab}{}^{cd}\right) \ , \tag{3.1b}$$

$$(\mathcal{P}\chi)_{\underline{\alpha}} \stackrel{*}{=} 0 \ , \quad \mathcal{P} \equiv \tfrac{1}{8}\left(I + \Psi\right) \ , \tag{3.1c}$$

$$F_{ab}^{(-)i} \stackrel{*}{=} 0 \ , \quad G_{ab}^{(-)} \stackrel{*}{=} 0 \ , \tag{3.1d}$$

$$\nabla_a \varphi \stackrel{*}{=} + \tfrac{1}{6\sqrt{2}} \phi_a{}^{bcd} H_{bcd} \ , \tag{3.1e}$$

$$R_{abcd}^{(-)} \equiv N_{cd}{}^{ef} R_{abef} \stackrel{*}{=} 0 \ , \tag{3.1f}$$

These constraints are extra associated with supersymmetric self-duality, in addition to (2.2). To clarify that these are such extra constraints, we use the symbol $\stackrel{*}{=}$. The convention for the octonionic structure constants $\psi_{abc}$ is like $\psi_{123} = \psi_{516} = \psi_{624} = \psi_{435} = \psi_{471} = \psi_{673} = \psi_{572} = +1$, $\phi_{abcd} \equiv (1/3!)\epsilon_{abcd}{}^{efg}\psi_{efg}$ [3][6]. The matrix $\mathcal{N}$ projects a spinor **8** into a **7** under $SO(7) \to G_2$. This is complementary to the matrix $\mathcal{P}$ projecting a **8** into a **1**. The anti-self-dual projector $N_{ab}{}^{cd}$ reduces an adjoint representation **21** into a **7** under $SO(7) \to G_2$. This is complementary to the self-dual component projector

$$P_{ab}{}^{cd} \equiv \tfrac{1}{3}\left(\delta_a{}^{[c}\delta_b{}^{d]} + \tfrac{1}{2}\phi_{ab}{}^{cd}\right) \ . \tag{3.2}$$

projecting a **21** into the adjoint representation **14** of $G_2$. The superscript $^{(-)}$ in (3.1) is the anti-self-dual component for the pair of indices $ab$, projected by the operator $N_{ab}{}^{cd}$. Even though we do not write explicitly, other constraints, such as $R_{\underline{\alpha}bcd}^{(-)} \equiv N_{cd}{}^{ef} R_{\underline{\alpha}bef} \stackrel{*}{=} 0$ or $R_{\underline{\alpha}\underline{\beta}\,cd}^{(-)} \stackrel{*}{=} 0$ will follow, as the necessary conditions of (3.1).

The consistency check of these constraints is performed by applying a spinorial derivative $\nabla_{\underline{\alpha}}$ on each of the extra constraints (3.1b) through (3.1f), with the aid of identities, such as $\psi_{abc}\phi^{abde} \equiv -4\psi_c{}^{de}$ [6]. First, the spinorial derivative acting on (3.1b) is shown to vanish:

$$0 \stackrel{?}{=} \nabla_{\underline{\gamma}} T_{ab}^{(-)\underline{\delta}} \stackrel{*}{=} -\epsilon_{ij} N_{ab}{}^{cd} (\mathcal{N}\tau_3)_{\underline{\gamma}}{}^{\underline{\delta}} F_{ce}{}^i F_d{}^{ej} - 2 N_{ab}{}^{cd} (\mathcal{N}\tau_i)_{\underline{\gamma}}{}^{\underline{\delta}} F_{ce}{}^i G_d{}^e \stackrel{*}{=} 0 \ , \tag{3.3}$$

due to the identity [6][3]

$$N_{ab}{}^{cd} P_{ce}{}^{fg} P_d{}^{ehk} \equiv 0 \ . \tag{3.4}$$

Second, the case of (3.1c) is also straightforward, owing to the useful identities:

$$(\mathcal{N}\gamma_a \mathcal{P})_{\alpha\beta} \equiv +i\delta_{a\alpha}\delta_{\beta 8} \ , \quad (\mathcal{N}\gamma_{ab}\mathcal{P})_{\alpha\beta} \equiv +\psi_{ab\alpha}\delta_{\beta 8} \ ,$$

$$(\mathcal{N}\gamma_c \mathcal{N})_{\alpha\beta} \equiv +i\psi_{c\alpha\beta} \ , \quad (\mathcal{N}\gamma_{cd}\mathcal{N})_{\alpha\beta} \equiv +\phi_{cd\alpha\beta} + \delta_{c[\alpha}\delta_{\beta]d} \ ,$$

$$(\mathcal{N}\gamma_{abc}\mathcal{N})_{\alpha\beta} \equiv -\tfrac{i}{2}\delta_{(\alpha|[a}\psi_{bc]|\beta)} - \tfrac{i}{2}\delta_{\beta(a}\psi_{bc]\alpha} + i\delta_{\alpha d}\delta_{d\beta}\psi_{abc} - \delta_{\alpha 8}\phi_{abc\beta} \ ,$$

$$(\mathcal{N}\gamma_{abc}\mathcal{P})_{\alpha\beta} \equiv \psi_{ab\alpha}\delta_{\beta 8} \ , \quad P_{cd}{}^{ef}\psi_{efg} \equiv 0 \ , \quad \psi_{abc}\phi^{abde} \equiv -4\psi_c{}^{de} \ . \tag{3.5}$$

Third, the case of (3.1d) is very simple, due to the self-duality $T^{(-)\gamma}_{ab} \stackrel{*}{=} 0$. Fourth, the spinorial derivative of (3.1e) needs a special care, because it needs the gravitino field equation (2.6a):

$$0 \stackrel{?}{=} \nabla_{\underline{\alpha}}\left(\nabla_a\varphi - \tfrac{1}{6\sqrt{2}}\phi_a{}^{bcd}H_{bcd}\right) \stackrel{*}{=} -\tfrac{1}{2\sqrt{2}}\tau_3\Big[ +\gamma^b T_{ab} - 2i\tau_3\nabla_a\chi + i(\gamma^{bc}\tau_3\chi)H_{abc}$$
$$+ \sqrt{2}(\gamma^b\tau^i\chi)F_{ab}{}^i + \sqrt{2}(\gamma^b\chi)G_{ab}\Big]_{\underline{\alpha}} \stackrel{.}{=} 0 \ . \quad (3.6)$$

Finally, the case of (3.1f) is

$$0 \stackrel{?}{=} \nabla_{\underline{\alpha}} R^{(-)}_{bcde} \equiv + \nabla_{[b} R^{(-)}_{\underline{\alpha}|c]de} + T_{\underline{\alpha}[b|}{}^f R^{(-)}_{f|c]de}$$
$$+ T_{\underline{\alpha}[c|}{}^{\underline{\eta}} R^{(-)}_{\underline{\eta}|c]de} - T_{bc}{}^f R^{(-)}_{\underline{\alpha} fde} + T]_{bc}{}^{\underline{\eta}} R^{(-)}_{\underline{\eta}\underline{\alpha}de} \stackrel{*}{=} 0 \ . \quad (3.7)$$

Here use is made of the $R$-Bianchi identity starting with $\nabla_{\underline{\alpha}} R_{bcde} + \cdots \equiv 0$, as well as other facts, such as $R^{(-)}_{\underline{\alpha}\underline{\beta}cd} \stackrel{*}{=} 0$ and $R^{(-)}_{\underline{\alpha}bcd} \stackrel{*}{=} 0$.

The on-shell degrees of freedom in this system are matched under supersymmetry in the following way: First, each of the three vectors $A_m{}^i$ and $B_m$ have only three degrees of freedom, so in total $3 \times 3 = 9$ degrees of freedom, due to their self-dualities (3.1d). The siebenbein $e_m{}^a$ has $(3\times 4)/2 - 1 = 5$ degrees of freedom, where $3$ is like that for the index $m$ of a self-dual vector $B_m$, and $(3\times 4)/2 = 6$ is for its symmetry, while the subtraction of unity is due to the traceless-ness. The relationship (3.1e) says simply that one degree of freedom by $\varphi$ is completely determined by the field strength $H_{abc}$. Namely, out of the sum $(5\times 4)/2 + 1 = 11$ of freedoms of $\varphi$ and $C_{mnr}$, only $11 - 1 = 10$ degrees of freedom remain. Therefore, all the bosonic fields have $9 + 5 + 10 = 24$ degrees of freedom. Now due to the self-duality (3.1b), the gravitino has only $(4 \times 4 \times 2)/2 = 16$ degrees of freedom, where the number $2^{[7/2]-1} = 4$ is for a Majorana spinor in 7D, the additional doubling is due to $N=2$ indices, and $7 - 3 = 4$ is for the index $m$ for $\psi_m$, while the division by $2$ is due to the self-duality (3.1b). Similarly, $\chi$ has $4 \times 2 = 8$ degrees of freedom. In total, there are $16 + 8 = 24$ degrees of freedom, matching those for the bosonic fields, as $24 + 24$.

As has been mentioned, the absence of the $^{(-)}$ components implies the absence of the $\mathbf{7}$ in $\mathbf{21} = \mathbf{14} + \mathbf{7}$ under $SO(7) \to G_2$, namely, these indices are reduced into the adjoint representation $\mathbf{14}$ of the reduced holonomy $G_2$. This is also reflected into the supersymmetry transformation of the Lorentz spinor connection $\omega_{bcd}$ constructed [14] from (2.5d) as

$$\delta_Q \omega_{bcd} = +i(\overline{\epsilon}\gamma_b\tau_3 T_{cd}) \ , \quad (3.8)$$

agreeing with the self-duality of the last two indices on $\omega_{bcd}$:

$$\omega_{bcd} \stackrel{*}{=} +\tfrac{1}{2}\phi_{cd}{}^{ef}\omega_{bef} \ , \quad (3.9)$$

thanks to the self-duality $T_{cd}{}^\delta = +(1/2)\phi_{cd}{}^{ef}T_{ef}{}^\delta$. As in the corresponding case in 8D [9], our BFFC has been chosen to be compatible with such a requirement. Exactly as in 8D [9], any other constraint set away from the BFFC will cause some problem in the component transformation rule $\delta_Q \omega_{abc}$, not consistent with the self-duality in the last two indices. Since this aspect is just parallel to the 8D case [9], we skip the details here.

The gravitational superfield equation (2.6b) is also consistent with the self-duality of the Riemann supercurvature. To be more specific, we have [9]

$$R_{ac} = +\delta^{bd}R_{abcd} \stackrel{*}{=} +\tfrac{1}{2}\phi_c{}^{def}R_{adef} = +\tfrac{1}{12}\phi_c{}^{def}R_{a[def]}$$
$$= +\tfrac{1}{12}\phi_c{}^{def}\Big(-4\nabla_a H_{def} - \tfrac{1}{2}F_{[ad}{}^i F_{ef]}{}^i + \tfrac{1}{2}G_{[ad}G_{ef]}\Big) \stackrel{*}{=} -2\sqrt{2}\nabla_a\nabla_b\varphi \ . \quad (3.10)$$

As in the analogous case in 8D [9], this is a modification of the usual Ricci flatness condition in the torsion-full space with the $G_2$ holonomy [3].

Compared with self-dual supergravity in 8D with reduced holonomy $Spin(7)$ [9], there are similarities and differences. The similarities are such as the self-dual Riemann tensor, or the maximal holonomy $SO(7)$ reduced to $G_2$, as our first desirable goals. Other technical similarities are such as the Ricci tensor with the second derivative of the dilaton as in (3.10), with a structure similar to 8D [9]. One difference is that the asymmetry based on chirality played an important role for the supersymmetric self-duality in 8D. For example, in 8D the spinor charge with *positive* chirality $\nabla_\alpha$ was constrained to **1** as **8** → **7** + **1** under $SO(8) \to Spin(7)$, while the gravitino for negative chirality was truncated $T_{ab}{}^\gamma \stackrel{*}{=} 0$ [9]. In this sense, the self-dual supergravity had $N = (1,0)$ supersymmetry in 8D. In 7D, we do not have such 'asymmetry' depending on the spinor charges $\nabla_{\alpha 1}$ and $\nabla_{\alpha 2}$, but the condition $\nabla_{\underline{\alpha}} \stackrel{*}{=} (\mathcal{N}\nabla)_{\underline{\alpha}}$ is common to both charges. In 7D, since both spinor charges are present, and in that sense we still have $N = 2$ supersymmetry, but it is a 'restricted' one.

## 4. Topological $N = 2$ Self-Dual Supergravity in 7D with Reduced Holonomy $G_2$

As we have promised in the Introduction, we next present an alternative 'topological' self-dual supergravity in 7D. The difference from the last section is that both of the spinor charges are now in the **1** of $G_2$, instead of **7**.

After trials and errors, we found such a set of constraints for supersymmetric generalized self-duality, as

$$\nabla_{\underline{\alpha}} \stackrel{*}{=} (\mathcal{P}\nabla)_{\underline{\alpha}} \ , \quad (4.1a)$$

$$T_{ab}^{(-)\underline{\gamma}} \equiv N_{ab}{}^{cd}T_{cd}{}^{\underline{\gamma}} \stackrel{*}{=} 0 \ , \quad (4.1b)$$

$$\chi_{\underline{\alpha}} \overset{*}{=} 0 ~, \tag{4.1c}$$

$$F^{(-)i}_{ab} \overset{*}{=} 0 ~, \quad G^{(-)}_{ab} \overset{*}{=} 0 ~, \tag{4.1d}$$

$$\varphi \overset{*}{=} 0 ~, \tag{4.1e}$$

$$\phi_a{}^{bcd} H_{bcd} \overset{*}{=} 0 ~, \quad \psi^{abc} H_{abc} \overset{*}{=} 0 ~, \tag{4.1f}$$

$$R^{(-)}_{abcd} \equiv N_{cd}{}^{ef} R_{abef} \overset{*}{=} 0 ~, \tag{4.1g}$$

These constraints are extra, in addition to the original BFFC set (2.2). In eq. (4.1g), the anti-self-duality symbol refers only to the last pair $cd$, but *not* the first one.

The consistency of this set of constraints for supersymmetric generalized self-duality can be confirmed, by applying fermionic derivatives on the constraints (4.1b) - (4.1g). After applying these derivatives, we can use these constrains again, in order to see whether they vanish consistently. A typical example is on (4.1c):

$$0 \overset{?}{=} \nabla_{\underline{\alpha}1} \chi_{\underline{\beta}1} \overset{*}{=} (\mathcal{P}\nabla)_{\alpha 1} \chi_{\beta 1}$$
$$\overset{*}{=} -\tfrac{i}{4\sqrt{2}}(\mathcal{P}\gamma^{cd}_{(-)})_{\alpha\beta}(F_{cd}{}^2 - G_{cd}) + \tfrac{i}{8}(\mathcal{P}\gamma^c)_{\alpha\beta}(\overline{\chi}_1 \gamma_c \chi_1) - \tfrac{i}{16}(\mathcal{P}\gamma^{cd})_{\alpha\beta}(\overline{\chi}_1 \gamma_{cd} \chi_1) \overset{*}{=} 0 ~, \tag{4.2}$$

where we have used (4.1c), (4.1d) and the identity $\mathcal{P}\gamma^{cd} \equiv \mathcal{P}\gamma^{cd}_{(-)}$. The subtlety arises with the fermionic derivative on the constraint (4.1b), because we need a peculiar lemma (3.4) for the $FG$-term $N_{ab}{}^{cd} F^{(+)i}_{ce} G^{(+)}_{de}$. As for the constraint (4.1g), we use the $R$-Bianchi identity $\nabla_{\underline{\alpha}} R_{bcde} + \cdots \equiv 0$, as in (3.7).

Since our fermionic derivatives $\nabla_{\underline{\alpha}}$ have only the singlet component $\mathbf{1}$ in $\mathbf{8} = \mathbf{7} + \mathbf{1}$ under $SO(7) \to G_2$, the usual commutator $\{\nabla_{\underline{\alpha}}, \nabla_{\underline{\beta}}\}$ vanishes, as the identity $\mathcal{P}\gamma^c \mathcal{P} \equiv 0$ also shows. This implies that this system of self-dual supergravity is 'topological', like nilpotent BRST symmetry. As a matter of fact, this is also consistent with the result in topological quantum field theory of self-dual (super)gravity [10]. The advantage of our formulation, however, is the usage of BFFC that drastically simplified the whole computation, compared with component formulation [10], where higher-order terms are considerably involved. We also see that the non-self-dual supergravity can be recovered by releasing the condition $\nabla_{\underline{\alpha}} = (\mathcal{P}\nabla)_{\underline{\alpha}}$, much like the link between the BRST formulation and supergravity discussed in [10].

Due to $R^{(-)}_{\underline{\alpha}bcd} \overset{*}{=} 0$, the self-duality $\omega^{(-)}_{bcd} \overset{*}{=} 0$ is also consistent with supersymmetry in this system, as in (3.8). Similarly, the Ricci-flatness of this system resembles (3.10):

$$R_{ac} = +\delta^{bd} R_{abcd} \overset{*}{=} +\tfrac{1}{2} \phi_c{}^{def} R_{adef} = +\tfrac{1}{12} \phi_c{}^{def} R_{a[def]}$$
$$= +\tfrac{1}{12} \phi_c{}^{def} \left( -4\nabla_a H_{def} - \tfrac{1}{2} F_{[ad}{}^i F_{ef]}{}^i + \tfrac{1}{2} G_{[ad} G_{ef]} \right) \overset{*}{=} 0 ~. \tag{4.3}$$

Compared with (3.10), or with the analogous case in 8D [9], the Ricci-flatness of this system with the $G_2$ holonomy [3] is not modified.[3] These results imply that the manifold we are dealing with in this formulation has definitely non-trivial $G_2$ holonomy, but supersymmetry is 'topological' [10], instead of the usual one generating translations.

The on-shell degrees of freedom are matched under supersymmetry as follows. First, the three self-dual vectors $A_m{}^i$ and $B_m$ have $3 \times 3 = 9$ degrees of freedom. Next, $C_{mn}$ has $(5 \times 4)/2 - 7 - 1 = 2$ degrees of freedom due to the $7+1$ conditions (4.1f). The siebenbein has $(3 \times 4)/2 - 1 = 5$ degrees of freedom as in the last section, while the dilaton has no freedom. The gravitino has $(4 \times 8)/2 = 16$ degrees of freedom, due to the self-duality (4.1b). Finally, $\chi_\alpha$ has zero degree of freedom. In total, the bosons and fermions have the same 16 degrees of freedom, as $16+16$ in this system.

Compared with the first formulation in the last section, there are similarities and differences. The similarities are such as the self-dual Riemann tensor and the reduced holonomy $G_2 \subset SO(7)$. The difference is that the dilaton and dilatino superfields are constrained to vanish, while the superfield strength $H_{abc}$ is subject to the constraint (4.1f) without the dilaton. Moreover, the spinorial derivative $\nabla_\alpha$ is subject to (4.1a). This is consistent with the Killing spinor condition for the singlet spinor $\mathbf{1}$ under the reduction of the holonomy $SO(7) \to G_2$ for the compactification of M-theory into 4D with chiral fermions [3][4]. It is this fundamental aspect that is reflected into the conditions, such as the vanishing of the dilatino (4.1c) and dilaton (4.1e), or the condition on $H_{bcd}$ (4.1f) without $\varphi$. In other words, this self-duality in 7D is more restrictive than the corresponding 8D case or the first formulation in the last section, with spinor charges realized only as singlets under the holonomy $G_2$.

We mention the possible 'twisted' version of $N=2$ supersymmetries, i.e., the hybrid of two supergravity theories in this paper, such as the twisting $\nabla_{\alpha 1} \stackrel{*}{=} (\mathcal{P}\nabla)_{\alpha 1}$ and $\nabla_{\alpha 2} \stackrel{*}{=} (\mathcal{N}\nabla)_{\alpha 2}$. We have tried to formulate this version, so far with no success. There are several technical obstructions according to our trials. First, we can no longer put both $\chi_{\alpha 1}$ and $\chi_{\alpha 2}$ to zero, because of non-vanishing term $\psi_{(\alpha}{}^{de} H_{\beta)de}$ arising in $\nabla_{\alpha 2} \chi_{\beta 1}$. This is also related to the non-vanishing of the combination $(\mathcal{N}\gamma^{cde})_{\alpha\beta} H_{cde}$. Second, even if we allow only $\chi_{\alpha 2}$ to be zero, the condition $\nabla_a \varphi \stackrel{*}{=} -(1/6\sqrt{2})\phi_a{}^{bcd} H_{bcd}$ is not consistent with its spinorial derivative, which requires the *opposite* sign between these two terms. There does not seem to be any way out to avoid these problems. Even though the result for global supersymmetry [16] indicate more naturally a twisted $N=2$ supergravity in 7D, we do not know any conceptual reason for this obstruction at the present time.

---

[3]The non-vanishing torsion $T_{ab}{}^c \equiv +2H_{ab}{}^c$ does not eventually affect the Ricci-flatness in this system, due to $\varphi \stackrel{*}{=} 0$.

We repeat that our formulation is based on the combination of the peculiar feature of the octonionic structure constants $\psi_{abc}$ leading to generalized self-duality, and the usage of BFFC constraints, all closely related to each other consistently in superspace.

## 5. Concluding Remarks

In this paper, we have presented two formulations of self-dual supergravity in Euclidean 7D with the reduced holonomy $G_2$. The first formulation has both of the $N = 2$ spinor charges as the **7** of $G_2$, while the second one has both spinor charges as in **1** of $G_2$. The second formulation is more closely related to compactifications of 11D M-theory into 4D, with singlet Killing spinors [2][3][4]. These formulations have certain differences from as well as similarities to the corresponding self-dual supergravity in 8D [9].

The similarities are such as the crucial role played by octonion structure constants, the importance of the special set BFFC for superspace constraints, or the self-dual Riemann tensor with the reduced holonomy $G_2 \subset SO(7)$. Or the simplification by superspace formulation itself is already the common feature in these dimensions, because the component formulations [10] will get more involved for fixing higher-order terms involving fermions. The most important difference is that the surviving supersymmetries invariant under the reduced holonomy $G_2$ required from compactifications [3][4], impose rather strong conditions on the fermionic derivatives, such as $\nabla_{\underline{\alpha}} \stackrel{*}{=} (\mathcal{P}\nabla)_{\underline{\alpha}}$ in the second formulation.

The analysis of globally supersymmetric self-dual theories in diverse dimensions [17] indicates that $N = 2$ supersymmetry is minimally needed in 7D. Our result in this paper is also consistent with this general conclusion that at least $N = 2$ supersymmetry is needed to maintain the desired self-duality. Our recent component formulation for self-dual supersymmetric Yang-Mills theory in 7D [16] is also consistent with this conclusion.

Another interesting feature of self-dual supergravity in 7D we have found is as follows: In our previous paper [9], we have mentioned the possible usage of dimensional reduction of self-dual supergravity in 8D into lower dimensions, including 7D. However, there is one caveat for this statement. As we have seen in this paper, such a dimensional reduction does not work in a simple way, but needs a special care. This is because of the different aspects of even vs. odd dimensions associated with fermions [17]. For example, we saw in [9] that the $N = (1, 0)$ chiral supersymmetry was crucial for the compatibility between self-duality and supersymmetry in 8D, while extended $N = 2$ supersymmetry is minimally required for self-duality in 7D. Due to such different features in 8D and 7D, a simple dimensional reduction does not work smoothly for getting self-dual supergravity in 7D. As a matter of fact, we have seen a similar situation already in global supersymmetry between 8D and 7D [16][17]. In fact, when we need supersymmetric self-duality in 7D, a naïve simple dimensional reduction from 8D does not give a clue for the necessity of $N = 2$ supersymmetry in 7D.

To our knowledge, our present paper gives the first complete formulations of self-dual supergravity in 7D including all higher-order terms, before quantization. We have presented two different versions of $N = 2$ self-dual supergravity, which will be of great importance for future studies of such self-dualities in 7D, in particular, explicitly formulated in superspace. Our second superspace formulation for topological supergravity in 7D also gives an important 'bridge' between topological (super)gravities [10] and conventional superspace formulation which was originally constructed for the conventional local supersymmetry generating translation. As a by-product, we have found it possible to formulate the $N = 2$ self-dual supergravity in 7D, with both spinor charges in the **7** of $G_2$, that was *not* predicted by M-theory compactifications [1][3][4].

In our supergravity formulations, the maximal holonomy $SO(7)$ in 7D is reduced into $G_2$ consistently with local supersymmetry. The compatibility of supergravity with reduced holonomies is analogous to certain supergravity formulations with no manifest Lorentz covariances in higher dimensions $D \geq 12$ [18]. To put it differently, there is accumulating evidence that reduced or non-manifest holonomies become more and more important in higher-dimensional supergravity theories in $D \geq 4$.

We are grateful to M. Duff for helpful discussions of surviving supersymmetries in compactifications.